\documentclass[11pt]{article}
\usepackage[dvips]{epsfig}
\usepackage[T1]{fontenc}
\usepackage[latin1]{inputenc}
\usepackage{graphicx}
\usepackage[english]{babel}
\usepackage{amsmath}
\usepackage{amssymb}
\usepackage{amsfonts}
\usepackage[T1]{fontenc}
\usepackage{color}
\usepackage{babel}
\usepackage{verbatim}
\usepackage[unicode=true,pdfusetitle,bookmarks=true,bookmarksnumbered=false,bookmarksopen=false,
 breaklinks=false,pdfborder={0 0 1},backref=false,colorlinks=true]{hyperref}
\hypersetup{linkcolor=blue,citecolor=blue}
\makeatletter
\usepackage[dvips]{epsfig}
\usepackage[T1]{fontenc}
\usepackage{color}
\usepackage{pdflscape}
\usepackage{cite}

\begin{document}

\title{\bf Geometric Perfect Fluids and Dark Side of the Universe}

\author{Metin G{\" u}rses $^a$\thanks{%
email: gurses@fen.bilkent.edu.tr},~Yaghoub Heydarzade $^a$\thanks{%
email: yheydarzade@bilkent.edu.tr}, and \c{C}etin \c{S}ent{\" u}rk $^b$\thanks{%
email: csenturk@thk.edu.tr}\\\\
\\{\small $^a$Department of Mathematics, Faculty of Sciences, Bilkent University, 06800 Ankara, Turkey}\\
{\small $^b$Department of Aeronautical Engineering,
University of Turkish Aeronautical Association, 06790 Ankara, Turkey}}
\date{\today}

\maketitle
\begin{abstract}
Recently we showed that in FLRW cosmology, the contribution from higher curvature terms in any generic metric gravity theory to the energy-momentum tensor is of the perfect fluid form. Such a geometric perfect fluid can be interpreted as a fluid remaining from the beginning of the universe where the string theory is thought to be effective. Just a short time after the beginning of the Universe, it is known that the Einstein-Hilbert action is assumed to be modified by adding all possible curvature invariants. We propose that the observed late-time accelerating expansion of the Universe can be solely driven by this geometric fluid. To support our claim, we specifically study the quadratic gravity field equations in $D$-dimensions. We show that the field equations of this theory for the FLRW metric possess a geometric perfect fluid source containing two critical parameters $\sigma_1$ and $\sigma_2$. To analyze this theory concerning its parameter space $(\sigma_1, \sigma_2)$, we obtain the general second-order nonlinear differential equation governing the late-time dynamics of the deceleration parameter $q$.  Hence using some present-day cosmological data as our initial conditions, our findings for the $\sigma_2=0$ case are as follows: $ (i)$  In order to have a positive energy density for the geometric fluid $\rho_g$, the parameter $\sigma_1$ must be negative for all dimensions up to $D = 11$, $(ii)$ For a suitable choice of $\sigma_1$, the deceleration parameter experiences signature changes in the past and future, and in the meantime it lies within a negative range which means that the current observed accelerated expansion phase of the Universe can be driven solely by the curvature of the spacetime, $(iii)$ $q$ experiences a signature change and  as the dimension $D$ of spacetime increases, this signature change happens at earlier and later times, in the past and future, respectively.
\end{abstract}
\maketitle

\newpage
\tableofcontents

\section{Introduction}

Although General Relativity (GR) has been immensely successful in explaining a wide range of gravitational phenomena, there are certain observations that have motivated researchers to consider modifications to the theory. Two of these motivations are: (i) Theoretical consistency of GR: Modifications to gravity theories can arise from attempts to reconcile GR with other fundamental theories, such as quantum gravity or string theory. GR is assumed to be  a low-energy approximation of a more complete
theory where the effective action  includes higher-curvature terms in addition to the usual Einstein-Hilbert term. Hence, modified gravity can be seen as an exploration of how gravity might behave at very low or high energy
limits where the effects of quantum physics become significant. See for instance \cite{damour, hor, deser, teit, rovelli}, and (ii)
Dark matter and dark energy: The need for dark matter and dark energy to explain the observed motion of galaxies and the accelerated expansion of the universe, respectively, has led to questions about whether our understanding of gravity is complete. Modified gravity theories also seek to address these phenomena without introducing the need for dark matter or dark energy
\cite{jacob, laur, bohmer, riazi, nojiri1, brane1, brane2 ,zukin, koyama, Car-etal, harada, odint, Noj2010, Noj2017 }.

Higher order curvature corrections to Einstein's field equations have  been considered by many authors
\cite{wand,  sch, cap, bar, noji0, DBD, PDG, lov, casal, noji2001}.
Recently we showed that, in Friedman-Lemaitre-Robertson-Walker (FLRW) cosmology,  the contribution of higher curvature terms in any generic theory of gravity to the energy momentum tensor is of the perfect fluid form \cite{gur1}. This is the reason that some authors have observed this fact \cite{chen, Man1, Man2} and verified it in some particular modified gravity theories such as $f(R)$ gravity \cite{Capo2006,fr}, Gauss-Bonnet
gravity \cite{capoz2}, and other higher order gravities \cite{higher1, higher2}.
In \cite{gur1},  the cases of general Lovelock and $F(R, G)$ theories
are given as examples.

The FLRW metric is the most known and most studied metric in general relativity. This metric is mainly used to describe the Universe as a homogeneous, isotropic
fluid distribution \cite{wein}.  It is  known that
FLRW cosmology in Einstein's theory is not sufficient to explain the accelerating expansion of the universe. To explain this phenomena, staying in Einstein theory it is claimed that, in addition to the known ordinary matter distribution, the Universe should contain a dark component
driving the accelerating expansion,
the so called Dark Energy. If we consider the low energy limit of the string theory, then the Einstein equations are modified by adding
all possible curvature invariants. We call such a theory as generic gravity where the action takes the form

\begin{equation}\label{genact}
I=\int\, d^{D}\,x\, \sqrt{-g}\,\left [\frac{1}{\kappa} \left(R-2\Lambda \right)+
\mathcal{F}(g,\,\mbox{Riem},\,\nabla \mbox{Riem},\, \nabla \nabla \mbox{Riem}, \dots)+ \mathcal{L}_{M} \right],
\end{equation}
 here ${\cal F}$ is a function of  all combinations of the metric tensor, curvature tensor, and the covariant derivatives of the curvature tensor of any order. Then the
field equations take the form
\begin{equation}\label{FEs}
\frac{1}{\kappa}\left(G_{\mu \nu}+ \Lambda g_{\mu \nu}\right)+{\cal E}_{\mu \nu}=T^{M}_{\mu \nu},
\end{equation}
where $G_{\mu \nu}$ is  the Einstein tensor, $\Lambda$ is the cosmological constant, $T^{M}_{\mu \nu}$ is the matter energy momentum tensor of perfect fluid distribution
(with the energy
density $\rho$ and pressure $p$), and $\mathcal{E}_{\mu \nu}$ resulting from the higher order curvature terms is
any second rank tensor obtained from the metric tensor, Riemann tensor,  Ricci tensor, Ricci scalar and their covariant derivatives at any order.
In \cite{gur1}, we have shown that $\mathcal{E}_{\mu \nu}$  can be written as
the combination of the metric tensor $g_{\mu \nu}$ and $u_{\mu} u_{\nu}$; that is,
\begin{equation}\label{E}
{\cal E}_{\mu \nu}=A g_{\mu \nu}+B u_{\mu} u_{\nu},
\end{equation}
where $A$ and $B$ are functions of scale factor $a(t)$ and its derivatives with respect to time $t$. This implies that in the Einstein field equations, in addition to the matter fluid
energy momentum tensor $T^{M}_{\mu \nu}$, there exists another fluid distribution which we call it as geometric fluid distribution
$T^{g}_{\mu \nu}$ with the energy density $\rho_{g}=A-B$ and pressure $p_{g}=-A$. Here, we adopt the idea that the source of the dark matter/energy is the geometrical fluid distribution. Hence we conjecture that higher curvature modifications of the Einstein theory are complete in the sense that all cosmological observations can be explained by choosing appropriate modified theories studied by several authors in \cite{wand,  sch, cap, bar, noji0, DBD, PDG, lov, casal, noji2001}.
To support our claim, we specifically study the quadratic gravity field equations in $D$-dimensions. We show that the field equations of this theory for the FLRW metric possess a geometric perfect fluid source containing two critical parameters $\sigma_1$ and $\sigma_2$. To analyze this theory concerning its parameter space $(\sigma_1, \sigma_2)$, we obtain the general second-order nonlinear differential equation governing the late-time dynamics of the deceleration parameter $q$.  Hence using some present-day cosmological data as our initial conditions, our findings for the $\sigma_2=0$ case are as follows:  $ (i)$  In order
to have a positive energy density for the geometric fluid $\rho_g$, the parameter $\sigma_1$ must be negative for all dimensions up to $D = 11$, $(ii)$ For a suitable choice of $\sigma_1$, the deceleration parameter
experiences signature changes in the past and future, and in the meantime
it lies within a negative range which means that the current observed accelerated expansion phase of the Universe can be driven solely by the curvature of the spacetime,
$(iii)$ $q$ experiences a signature change and  as the dimension $D$ of spacetime increases, this signature change happens at earlier and later times, in the past and future respectively,
$(iv)$ The geometric equation of state parameter  $w_g$ is negative for all integers $4\leq D\leq 10$, specifically, for $D=4$, $w_g=-0.85$.
For $\sigma_1=0$ (Critical Gravity), we find that there are two cases both representing a possibility of having an accelerating expansion. Furthermore, we present some particular cosmological solutions in quadratic gravity depending upon the parameters $\sigma_{1}$ and $\sigma_{2}$.

It is well-known that linearized versions of most higher
time-derivative theories suffer from the Ostragradsky instability (see, for instance, \cite{wood1,wood2}). One way to avoid this instability is to consider such theories as low
energy approximations to a more fundamental theory, such as string
theory. Namely, at the scales where negative norm states appear, the
theory is expected to be replaced by a better-behaved UV theory. Another
possibility is Weinberg's asymptotically safe gravity \cite{wein0} where there are
infinitely many powers of curvature and the negative norm states
appear only in the truncated, perturbative version of the theory and disappear in the non-perturbative formulation where all
the coupling constants, i.e. all higher derivative curvature terms, are
taken into account. In \cite{kunz}, it is also remarked that Ostrogradskian ghosts in higher-derivative gravity theories
(generic gravity theories) are
only apparent when one truncates the infinite series of curvature invariants,
and hence these ghosts can be removed by means of a suitable boundary condition. Furthermore,  the absence of the Ostrogradsky instability manifests itself in theories with multiple fields; for example, in \cite{rham}, the authors discuss that, in the Extended-Scalar-Tensor class of theories for which the tensors are well-behaved and the scalar is free from gradient or ghost instabilities on FLRW spacetimes, one recovers the Horndeski theory up to field redefinitions. The general theorem introduced  in the present paper addresses  all
theories
that might be the low energy limit of the string theory, where the Einstein-Hilbert action and
hence the field equations
are modified by adding all possible curvature invariants. Based on this reason,
we call such a theory as generic gravity.  All our treatments such as proving our main theorem and all other derivations are non-perturbative. There
is no truncation, and hence  the Ostragradsky instabilities are not in the scope
of the present work.

In the next section, we summarize our theorem for the $D-$dimensional FLRW metrics. For illustration, we shall study the quadratic gravity theory in detail.

\section{Generic Gravity Field Equations in Perfect Fluid Form}
Using the covariant decomposition, one
can write the $D$-dimensional FLRW metric as
\begin{equation}\label{metric}
g_{\mu\nu}=-u_\mu\, u_\nu+a^2\,h_{\mu\nu},
\end{equation}
where $\mu,\,\nu=0...D-1$, $a=a(t),~u_\mu=\delta^0_\mu$, and $h_{\mu\nu}$
reads as
\begin{equation}
h_{\mu\nu}=\begin{pmatrix}
0 & 0 & \hdots & 0 \\
0 &  &  &  \\
\vdots &  & h_{ij} &  \\
0 &  &  &  \\
\end{pmatrix},
\end{equation}
where $h_{ij}=h_{ij}(x^{a})$  is the metric of the spatial section of the spacetime possessing the constant curvature $k$, and $i,j=1,...,D-1$.
One notes that
\begin{eqnarray}
&& u^\mu \,h_{\mu\nu}=u_\mu\, h^{\mu\nu}=0,\nonumber\\
&&h^{\mu}_{\alpha}= h^{\mu\alpha}\,h_{\alpha\nu}=\delta^{\mu}_{\nu}
+u^\mu \,u_\nu.
\end{eqnarray}
The corresponding Christoffel symbols can be obtained as
\begin{equation}\label{christ}
\Gamma^\mu_{\alpha \beta}=\gamma^\mu_{\alpha\beta}-a\,\dot a\, u^\mu\, h_{\alpha \beta}+H\,\left( 2u_\alpha\, u^\mu\, u_\beta +u_\beta\, \delta^\mu_\alpha +u_\alpha\,
\delta^\mu_\beta \right),
\end{equation}
where the  dot sign represents the derivative with respect to time $t$, $H=\dot a/a$ and $\gamma^\mu_{\alpha\beta}$ is defined as
\begin{equation}
\gamma^\mu_{\alpha\beta}=\frac{1}{2}a^2\, h^{\mu\gamma}\,\left(h_{\gamma \alpha,\beta}+h_{\gamma \beta,\alpha} -h_{\alpha \beta,\gamma} \right).
\end{equation}
The  Riemann curvature tensor can be written in the following
linear form in terms of the metric  $g_{\alpha\gamma}$ and the four vectors $u_\alpha$  and $u^\mu$
\begin{equation}\label{riemann2}
R^{\mu}_{\alpha\beta\gamma}=\left[  \delta^\mu_\beta g_{\alpha\gamma}-\delta^\mu_\gamma g_{\alpha\beta}\right]\rho_1+\left[u^{\mu}\left(g_{\alpha\gamma} u_\beta -
g_{\alpha\beta} u_\gamma \right)-u_\alpha\,\left(\delta^\mu_\gamma u_\beta -  \delta^\mu_\beta u_\gamma \right)\right]\rho_2,
\end{equation}
where $\rho_1$ and $\rho_2$ are defined as
\begin{eqnarray}
&&\rho_1=H^2 +\frac{k}{a^2},\label{rho1}\\
&&\rho_2=H^2 +\frac{k}{a^2}-\frac{\ddot a}{a}=\dot{H}+\frac{k}{a^2}.\label{rho2}
\end{eqnarray}
 The contractions of the Riemann tensor (\ref{riemann2}) gives the Ricci tensor
and Ricci scalar, respectively, as
\begin{eqnarray}\label{ricci}
&&R_{\mu\nu}=P g_{\mu\nu}+Qu_\mu u_\nu, \nonumber\\
&&R=DP-Q,
\end{eqnarray}
where
\begin{eqnarray}
P&=&(D-1) \rho_{1}-\rho_{2},\\
Q&=&(D-2) \rho_{2}.
\end{eqnarray}
One can also verify that the Weyl tensor
\begin{equation}\label{weyl}
C^{\mu}_{\alpha\beta\gamma}=R^{\mu}_{\alpha\beta\gamma}+\frac{1}{D-2}\left( \delta^\mu_\gamma R_{\alpha\beta}-\delta^\mu_\beta R_{\alpha\gamma}+ g_{\alpha\beta}R^\mu_\gamma -
g_{\alpha\gamma} R^\mu_\beta\right)+\frac{1}{(D-1)(D-2)}\left(  \delta^\mu_\beta g_{\alpha\gamma}-\delta^\mu_\gamma g_{\alpha\beta}\right)R
\end{equation}
vanishes for the metric (\ref{metric}). Since the conformal tensor is zero, the curvature tensor is expressed in terms of the Ricci tensor. This means that, for the FLRW
spacetime, the basic geometrical tensors are the metric and Ricci tensors. The covariant derivatives of the Ricci tensor are given by
\begin{eqnarray}
&&R_{\mu \nu;\alpha}=\dot P\,u_{\alpha}\, g_{\mu \nu}-Q H \left(u_{\nu}\, g_{\mu \alpha}+u_{\mu}\,g_{\nu \alpha}\right)+\left(\dot Q-2Q H\right) u_{\mu}\, u_{\nu}\,
u_{\alpha}, \\
&&{\square R}_{\mu \nu}=-\left[\ddot P+(D-1) H \dot P-2 Q H^2\right]\, g_{\mu \nu}+\left[2 D Q H^2-\ddot Q-(D-1) H \dot Q\right] u_{\mu} u_{\nu},\\
&&\square R=-D \ddot P-D(D-1) H \dot P+\ddot Q+(D-1) H \dot Q.
\end{eqnarray}
(NOTE: The above equations are from \cite{gur1} (Eq.(19)) where some typos are corrected.)\\

\vspace{0.2cm}
\noindent
The field equations of any generic gravity theory in D dimensions with the action, together with the matter fields,
\begin{equation}\label{action}
I=\int\, d^{D}\,x\, \sqrt{-g}\,\left [\frac{1}{\kappa} \left(R-2\Lambda \right)+
\mathcal{F}(g,\,\mbox{Riem},\,\nabla \mbox{Riem},\, \nabla \nabla \mbox{Riem}, \cdots)+ \mathcal{L}_{M} \right],
\end{equation}
 take the form
\begin{equation}\label{FEs}
\frac{1}{\kappa}(G_{\mu \nu}+ \Lambda g_{\mu \nu})+{\cal E}_{\mu \nu}=T^M_{\mu \nu},
\end{equation}
where $G_{\mu \nu}$ is  the Einstein tensor, $\Lambda$ is the cosmological constant, $T^M_{\mu \nu}$ is the energy momentum tensor coming from the matter fields denoted by $
\mathcal{L}_{M} $, and $\cal{E}_{\mu \nu}$ resulting from the higher order curvature terms contained in the function $\mathcal{F}$ is any second rank tensor obtained from the
metric tensor, Riemann tensor, Ricci tensor, and their covariant derivatives at any order. Hence we arrive at the following theorem \cite{gur1}:

\vspace{0.3cm}
\noindent
{\bf Theorem 1 }: {\it For the $D$-dimensional FLRW spacetimes given in (\ref{metric}), any second rank symmetric tensor obtained from the metric tensor,
Riemann tensor, Ricci tensor, and their covariant derivatives at any order becomes a combination of the metric tensor $g_{\mu \nu}$ and $u_{\mu} u_{\nu}$; that
is,
\begin{equation}\label{Emunu}
{\cal{E}}_{\mu \nu}=A g_{\mu \nu}+B u_{\mu} u_{\nu},
\end{equation}
where $A$ and $B$ are functions of the scale factor $a(t)$ and its time derivatives at any order and they depend on the underlying gravity theory.}

\vspace{0.3cm}
\noindent
And we have the following corollary of this theorem:

\vspace{0.3cm}
\noindent
{\bf Corollary 2}: {\it The field equations of any generic gravity theory given in (\ref{FEs}) take the form
\begin{equation}\label{FEs1}
\frac{1}{\kappa}(G_{\mu \nu}+ \Lambda g_{\mu \nu})=T^M_{\mu \nu}+T^g_{\mu \nu},
\end{equation}
where  $T^M_{\mu \nu}$ is the energy momentum tensor of perfect fluid distribution representing the baryonic matter fields,
\begin{equation}
T^M_{\mu \nu}=(\rho+p)u_\mu u_\nu+pg_{\mu\nu},
\end{equation}
with $\rho$ and $p$ being respectively the energy density and pressure of the fluid, and $T^g_{\mu \nu}$ is the tensor coming from the higher order curvature terms in
(\ref{action}) which is also in the perfect fluid form,
\begin{equation}
T^g_{\mu \nu}=(\rho_g+p_g)u_\mu u_\nu+p_gg_{\mu\nu},
\end{equation}
with $\rho_g=A-B$ and $p_g=-A$, due to (\ref{Emunu}). Hence defining an effective energy-momentum tensor as
\begin{equation}
T^{eff}_{\mu \nu}\equiv T^M_{\mu \nu}+T^g_{\mu \nu}=(\rho_{eff}+p_{eff})u_\mu u_\nu+p_{eff}g_{\mu\nu},
\end{equation}
the generic gravity field equations (\ref{FEs1}) for the FLRW spacetime with a perfect fluid source become
\begin{eqnarray}
&&\frac{1}{\kappa}\left[\frac{(D-1)(D-2)}{2}\rho_{1}-\Lambda\right]=\rho+A-B\equiv\rho_{eff},\label{rho}\\
&&-\frac{1}{\kappa}\left[\frac{(D-1)(D-2)}{2}\rho_{1}-(D-2)\rho_{2}-\Lambda\right]=p-A\equiv p_{eff}. \label{pres}
\end{eqnarray}
}
Thus the interpretations of the functions $A$ and $B$ appearing in the above formulation can be given as follows: the combination ``$A-B$'' is the energy density and ``$-A$'' is the pressure of a perfect fluid of purely geometric origin. The functions $A$ and $B$ differ in different modified theories. In each modification it is possible to arrange parameters of the theories to meet the observations. In particular, we shall analyze the quadratic gravity theory in $D$ dimensions in Section 4 and show that these purely geometric terms solely drive the late time accelerating expansion of the Universe consistently with the current observations.
In the next section, we shall give the cosmological parameters for generic gravity theories described by the action in (\ref{action}) in $D$ dimensions.

\section{Cosmological Parameters in Generic Gravity Theories}

Using (\ref{rho1}) and (\ref{rho}), one can obtain
\begin{eqnarray}\label{H2}
H^2 &=&\frac{2\kappa}{(D-1)(D-2)}\left[\rho+A-B+\frac{\Lambda}{\kappa}\right] -\frac{k}{a^2}\nonumber\\
&=&\frac{2\kappa}{(D-1)(D-2)}\sum_{i}\rho_i -\frac{k}{a^2},
\end{eqnarray}
where the label `$i$' denotes `$m$',~ `$r$',~ `$\Lambda$', or `$g$' representing matter, radiation,
cosmological constant (or dark energy), and ``dark geometric fluid'', respectively.
Defining the corresponding  dimensionless density parameter for each of the
mentioned components of the Universe as
\begin{equation}\label{omega}
\Omega_i=\frac{\rho_i}{\rho_{crit}} ~~~~\text{with}~~~~\rho_{crit}=\frac{(D-1)(D-2)H^2}{2\kappa},
\end{equation}
we can write (\ref{H2}) as
\begin{equation}\label{om1}
1=\Omega_m +\Omega_r +\Omega_\Lambda +\Omega_g +\Omega_k,
\end{equation}
where
\begin{equation}
\Omega_m=\frac{\rho_m}{\rho_{crit}},~~~~\Omega_r=\frac{\rho_r}{\rho_{crit}},
~~~~\Omega_\Lambda=\frac{\rho_\Lambda}{\rho_{crit}},~~~~\Omega_g=\frac{A-B}{\rho_{crit}},
~~~~\Omega_k=-\frac{k}{ a^2 H^2}.\label{omegas}
\end{equation}
and note that $\rho_\Lambda=\Lambda/\kappa$. The first observation here is the contribution
 of the  dark geometric fluid in determining the spatial curvature of the Universe.
One observes that
\begin{eqnarray}
&&\Omega_m +\Omega_r +\Omega_\Lambda +\Omega_g<1 ~ \Leftrightarrow\ \textnormal{open universe ($k=-1$)} \\
&&\Omega_m +\Omega_r +\Omega_\Lambda +\Omega_g=1 ~ \Leftrightarrow\ \textnormal{flat universe ($k=0$)} \\
&&\Omega_m +\Omega_r +\Omega_\Lambda +\Omega_g>1 ~ \Leftrightarrow\ \textnormal{closed universe ($k=1$)} .
\end{eqnarray}

On the other hand, using (\ref{rho1}), (\ref{rho2}), and (\ref{H2}), the equation (\ref{pres}) can be written equivalently as
\begin{equation}\label{press1}
-\frac{\ddot a}{aH^2}=\frac{\kappa}{(D-1)(D-2)H^2}\left[(D-3)\rho+(D-1)p-2A-B-2\frac{\Lambda}{\kappa}\right].
\end{equation}
Now defining the left hand side as the deceleration parameter $q$ and considering the barotropic equation of state $p_j=\omega_j \rho_j$, where $j=(m,r)$, for the matter
($\omega_m=0$) and the radiation ($\omega_r=1/3$), the equation (\ref{press1}) can be rewritten, with the help of (\ref{omega}), as
\begin{eqnarray}
q\equiv-\frac{\ddot a}{aH^2}&=&\frac{1}{2}\left[\sum_j[(D-3)+(D-1)w_j]\Omega_j-2\Omega_\Lambda-\Omega^*_g\right]\nonumber\\
&=&\frac{1}{2}\left[(D-3)\Omega_m+\frac{2}{3}(2D-5)\Omega_r-2\Omega_\Lambda-\Omega^*_g\right], \label{deccel}
\end{eqnarray}
where
\begin{equation}
\Omega^*_g=\frac{2A+B}{\rho_{crit}}.\label{omega*}
\end{equation}
Here we see that the geometric fluid has negative contribution to the deceleration (or, equivalently, positive contribution to the acceleration) of the Universe, if $2A+B>0$.
It can be observed that, setting $D=4$ and neglecting higher curvature modifications, one recovers
the deceleration parameter in GR as
\begin{equation}\label{}
q=\frac{1}{2}\left(\Omega_m+2\Omega_r-2\Omega_\Lambda \right).
\end{equation}

At this point, before proceeding further, it is appropriate to introduce the cosmological parameters
\begin{equation}
j(t)\equiv\frac{\dddot a}{aH^3},~~~~ s(t)\equiv\frac{\ddddot a}{aH^4},
\end{equation}
which are called ``jerk'' and ``snap'', respectively. These parameters, together with $H$ and $q$, are defined by expanding the scale factor in a Taylor series in the vicinity of
the current time $t_0$ \cite{visser}:
\begin{equation}
\frac{a(t)}{a_0}=1+H_0(t-t_0)-\frac{1}{2}q_0H_0^2(t-t_0)^2+\frac{1}{3!}j_0H_0^3(t-t_0)^3+\frac{1}{4!}s_0H_0^4(t-t_0)^4+O((t-t_0)^5),
\end{equation}
where $H_0$, $q_0$, $j_0$, and $s_0$ are the present-day values of the Hubble, deceleration, jerk, and snap parameters and they can be used to determine the evolutionary
behavior of the Universe.

\section{Quadratic Gravity and Criticality in $D$-Dimensions}

The action of the quadratic gravity theory \cite{GT,DT} is
\begin{eqnarray}
I & = & \int d^{D}x\,\sqrt{-g}\left[\frac{1}{\kappa}\left(R-2\Lambda\right)+\alpha R^{2}+\beta
R_{\mu\nu}^{^{2}}+\gamma\left(R_{\mu\nu\sigma\rho}^{2}-4R_{\mu\nu}^{2}+R^{2}\right)+\mathcal{L}_M\right],\label{eq:Quadratic_action}
\end{eqnarray}
giving the field equations
\begin{equation}
\frac{1}{\kappa}\left(R_{\mu\nu}-\frac{1}{2}g_{\mu\nu}R+\Lambda g_{\mu\nu}\right)+{\cal{E}}_{\mu\nu}=(\rho+p)\, u_{\mu}\,
u_{\nu}+p \, g_{\mu \nu},\label{fieldequations}
\end{equation}
where $\rho$ and $p$ are the energy density and pressure of the matter perfect fluid. Considering the FLRW metric \eqref{metric}, we find
\begin{eqnarray}
{\cal{E}}_{\mu \nu}&\equiv&2\alpha R\left(R_{\mu\nu}-\frac{1}{4}g_{\mu\nu}R\right)+\left(2\alpha+\beta\right)\left(g_{\mu\nu}\square-\nabla_{\mu}\nabla_{\nu}\right)R\nonumber \\
&&+2\gamma\left[RR_{\mu\nu}-2R_{\mu\sigma\nu\rho}R^{\sigma\rho}+R_{\mu\sigma\rho\tau}R_{\nu}^{\phantom{\nu}\sigma\rho\tau}-2R_{\mu\sigma}R_{\nu}^{\phantom{\nu}\sigma}-\frac{1}{4}g_{\mu\nu}\left(R_{\tau\lambda\sigma\rho}^{2}-4R_{\sigma\rho}^{2}+R^{2}\right)\right]\nonumber\\
&&+\beta\,\square\left(R_{\mu\nu}-\frac{1}{2}g_{\mu\nu}R\right)+2\beta\left(R_{\mu\sigma\nu\rho}-\frac{1}{4}g_{\mu\nu}R_{\sigma\rho}\right)R^{\sigma\rho}\nonumber\\ &=&
A g_{\mu \nu}+B u_{\mu} u_{\nu},\label{Emunu1}
\end{eqnarray}
where $A$ and $B$ are given by
\begin{eqnarray}
A&=&\frac{\alpha}{2}\left[ \left(D-1\right)\left(D\rho_1-2\rho_2\right) \right]
\left[-(D-1)(D-4) \rho_1 + 2(D-3)\rho_2\right]\nonumber\\
&&+(2\alpha+\beta)\left[  (D-1)(D-2)H\left(-D\dot \rho_1 +2\dot\rho_2\right) -D(D-1)\ddot \rho_1 +2(D-1)\ddot \rho_2 \right]\nonumber\\
&&+\frac{\gamma}{2} \left[ (D-2)(D-3)(D-4)\left(-(D-1)\rho_1^2 +4\rho_1 \rho_2   \right) \right]\nonumber\\
&&+\frac{\beta}{2}\left[  (D-2)\left[(D-1)(\ddot\rho_1 +(D-1)H\dot\rho_1)   \right]-2(\ddot\rho_2 +(D-1)H\dot\rho_2 -2H^2\rho_2)\right]\nonumber\\
  && +\frac{\beta}{2}\left[ -(D-1)(D-4)\left((D-1)\rho_1^2 +\rho_2^2\right)+4(D^2
  -5D +5)\rho_1\rho_2 \right],\\
 &&\nonumber\\
B&=&2\alpha\left[ \left(D-1\right)(D-2)\left(D\rho_1-2\rho_2\right)\rho_2 \right]\nonumber\\
&&+(2\alpha+\beta)\left[  (D-1)\left[H\left(-D\dot\rho_1 -2\dot\rho_2\right) -(D\ddot \rho_1 -2\ddot \rho_2)\right] \right]\nonumber\\
&&+2\gamma\left[ (D-2)(D-3)(D-4)\rho_1 \rho_2 \right]\nonumber\\
&&-\beta\left[(D-2)\left[\ddot\rho_2 +(D-1)H\dot\rho_2 -2DH^2\rho_2) \right]\right]\nonumber\\
  && +2\beta\left[(D-2)^2\rho_1\rho_2\right].
\end{eqnarray}
Hence, we can write (\ref{fieldequations}) as
\begin{eqnarray}
&&\frac{ 1}{\kappa }\left[\frac{1}{2} (D-2) (D-1) \left(\frac{k}{a^2}+\frac{\dot
a^2}{a^2}\right)-\Lambda \right]=\rho+\rho_{g_1} + \rho_{g_2}\equiv\rho_{eff},\label{reff}\\
&&-\frac{1}{\kappa }\left[\frac{(D-2)(D-3)}{2}  \left(\frac{k}{a^2}+\frac{\dot
a^2}{a^2}\right)+(D-2) \frac{\ddot a}{a}-\Lambda   \right]= p+ p_{g_1}+ p_{g_2}\equiv p_{eff},\label{peff}
\end{eqnarray}
where $\rho_{g_1}, \rho_{g_2}, p_{g_1}$, and $p_{g_2}$ have the geometric
origin given by
\begin{eqnarray}
&&\rho_{g_1} =-\frac{(D-1)\sigma_1}{2 D a^4} \Biggl\{k^2 (D-2)^2+\dot a^2 \left[2D (D-3)  a \ddot a-4 k (D-2)\right]
\nonumber\\
&&~~~~~~~~~~~~~~~~~~~~~~~~~~~~~~~~~~~~~~~~~~~~~~~~~~~~~~~~~~~-\left(D^2-4\right)\dot a^4-D a^2 \ddot
a^2+2 D a^2 \dot a \dddot a\Biggr\},\label{rg1}\\
&&\rho_{g_2} = -(D-1)\sigma_2\left(\frac{k}{a^2}+\frac{\dot
a^2}{a^2}\right)^2,\label{rg2}\\
&&p_{g_1} =-\frac{\sigma_1}{2Da^4}\Biggl\{(D-2) (D-5) (D+2) \dot a^4-k^2 (D-2)^2 (D-5)  +a \ddot
a \left(8 k (D-2)-3 D(D-3)a \ddot a\right)\nonumber\\
&&~~~~~~~~~+2 \dot a^2 \left[2 k (D-2) (D-5)-\left((D-9) D^2+12D+8\right) a \ddot a\right] -2 D \ddddot a a^3-4 (D-3)D \dddot a a^2 \dot a\Biggr\},\label{pg1}\\
&&p_{g_2} =\sigma_2\left(\frac{k}{a^2}+\frac{\dot
a^2}{a^2}\right) \left[(D-5) \left(\frac{k}{a^2}+\frac{\dot
a^2}{a^2}\right)+4 \frac{\ddot a}{a}\right],\label{pg2}
\end{eqnarray}
with
\begin{eqnarray}
&&\sigma_{1} = 4  (D-1)\alpha+D \beta,\label{s11}\\
&&\sigma_{2} = \frac{ (D-2) (D-4)}{2 D}\left[(D-1) (D-2) \alpha+ D(D-3)\gamma  \right]\label{s22}.
\end{eqnarray}
{\bf Remark 3}: When the parameters $\sigma_1$ and $\sigma_2$ vanish together, the geometric contributions given in (\ref{rg1})-(\ref{pg2}), resulting from the higher curvature terms in the action (\ref{fieldequations}), vanish identically and the equations (\ref{reff}) and (\ref{peff}) reduce to  the equations
in pure Einstein's gravity. These so-called critical points; i.e. $\sigma_1=0$ and $\sigma_2=0$, were identified and studied in higher curvature gravity theories first in four dimensions (where $\sigma_2$ identically vanishes) in \cite{LP} and later in $D$ dimensions in \cite{DT1}. In these works, it is shown that the linearized excitations around these critical points have vanishing energies and the mass and corresponding entropy of the usual Schwarzschild-AdS black hole solution turn out to be zero at criticality.

\vspace{0.2cm}
\noindent
{\bf Remark 4}: It can also be  observed that the above expressions are invariant under the scale transformations $ a \to \eta a$ and $k \to \eta^2 k$, where $\eta$ is a constant.

\vspace{0.2cm}
\noindent
{\bf Remark 5}: From the positiveness of $\rho_{g_2}$ in (\ref{rg2}) it follows that $\sigma_{2} <0$.

\vspace{0.2cm}
\noindent
From the equations (\ref{reff}) and (\ref{peff}), we can also write
\begin{equation}\label{add}
\frac{\ddot a}{a}=-\frac{\kappa}{(D-1)(D-2)}\left[(D-3)\rho_{eff}+(D-1)p_{eff}-2\frac{\Lambda}{\kappa}\right].
\end{equation}
Thus, when $\Lambda=0$, to have a universe expanding in an accelerating fashion ($\ddot{a}>0$), it must be that
\begin{equation}
w_{eff}=\frac{p_{eff}}{\rho_{eff}}<-\frac{D-3}{D-1}.\label{weff:cond}
\end{equation}
Specifically, for $D=4$, $w_{eff}<-1/3$. Using (\ref{reff}), along with the definition of the Hubble parameter, $H=\dot{a}/a$, one can also write (\ref{add}) as
\begin{equation}\label{Hdot}
\dot{H}=-\frac{\kappa}{D-2}(\rho_{eff}+ p_{eff})+\frac{k}{a^2},
\end{equation}
which is more convenient than (\ref{add}) because it does not involve the cosmological constant $\Lambda$ explicitly. This equation relates the acceleration of the Universe to its energy and momentum content. Together with the expressions (\ref{rg1})-(\ref{pg2}), the equation (\ref{Hdot}) becomes highly nonlinear and therefore it is not possible to give a compact analytical solution for $a(t)$. However, we can use (\ref{Hdot}) to investigate the late-time accelerated expansion of the Universe during which the higher curvature terms in the action (\ref{eq:Quadratic_action}) are assumed to be dominant. Indeed, recently it was observed that the Universe had entered into an accelerated expansion phase \cite{Riess-etal,Perl-etal} and a possible cause of this late time acceleration is the curvature of the spacetime itself related to the combination of the higher curvature terms that may appear in the action of a generic
gravity theory. This observation implies that the acceleration, defined by $\ddot{a}(t)$, of the Universe has changed its sign from negative to positive, or in other words, the deceleration parameter $q(t)$ has recently experienced a sign change from positive to negative. Therefore, for the purpose of investigating the late time acceleration or deceleration behavior of the Universe, we shall work with the deceleration parameter $q(t)$, instead of $a(t)$, and convert (\ref{Hdot}) as a differential equation for $q(t)$ which can be solved
numerically for given initial values of $q$ and its derivatives. Let us first write $\rho_{eff}$ and $p_{eff}$ in terms of the Hubble ($H$), deceleration ($q$),
jerk ($j$), and snap ($s$) parameters. From (\ref{rg1})-(\ref{pg2}),
\begin{eqnarray}
&&\rho_{eff}=\rho-\frac{\sigma_1(D-1)}{2 D} \Biggl[\frac{k^2(D-2)^2}{a^4}-4\frac{k}{a^2}(D-2)H^2\nonumber\\
&&~~~~~~~~~~~~~~~~~~~~~~~~~~~~~~~~~~~~~~~~~~~~-2D(D-3)qH^4-(D^2-4)H^4-Dq^2H^4+2DjH^4\Biggl] \nonumber\\
&&~~~~~~~~~~~~~-\sigma_2(D-1)\left(\frac{k}{a^2}+H^2\right)^2,\label{rhoeff}\\
&&p_{eff}=p-\frac{\sigma_1}{2 D} \Biggl[-\frac{k^2}{a^4}(D-2)^2(D-5)+4\frac{k}{a^2}(D-2)H^2[D-5-2q]+(D-2)(D+2)(D-5)H^4\nonumber\\
&&~~~~~~~~~~~~~~~~~~~~~~~~~~~~~~-3D(D-3)q^2H^4+2(D^3-9D^2+12D+8)qH^4-2DsH^4-4(D-3)DjH^4\Biggr],\nonumber\\
&&~~~~~~~~~~~~~+\sigma_2\left(\frac{k}{a^2}+H^2\right)\left[(D-5)\left(\frac{k}{a^2}+H^2\right)-4qH^2\right],\label{preseff}
\end{eqnarray}
where
\begin{equation}\label{js}
j=-\frac{\dot{q}}{H}+q(1+2q),~~~~s=-\frac{\ddot{q}}{H^2}+2\frac{\dot{q}}{H}(1+3q)-q(1+2q)(1+3q).
\end{equation}
In getting these expressions, use has been made of
\begin{equation}\label{q}
q=-\frac{\ddot{a}}{aH^2}=-\frac{\dot{H}}{H^2}-1.
\end{equation}
Now we shall set the ordinary matter and curvature to zero in the formulation, i.e. $\rho=0=p$ and $k=0$, to both comply with the observations and investigate the late-time acceleration of the
Universe resulting purely from the geometric terms related to curvature  of the spacetime. Then, (\ref{Hdot}) becomes an equation involving $H$, $q$, and the first and second
derivatives of $q$ with respect to time; that is,
\begin{eqnarray}\label{Hdot1}
&&\left[1+\frac{4H^2 \kappa \sigma_2} {D-2}\right](q+1)+\frac{\kappa \sigma_1 }{(D-2) D}\Biggl\{-2 H^2 \left(D^2-4\right)+H (D-7) D \dot{q}+D \ddot{q}\nonumber\\
&&~~~~~~~~~~~~~+H q \left[H \left(-6 D^2+15 D+8\right)-6 D \dot{q}\right]-4 H^2 (D-5) D q^2 + 6 H^2 D q^3\Biggr\}=0.
\end{eqnarray}
However, this form is not appropriate for solving the equation numerically. To obtain an appropriate form, we will change the time derivatives to derivatives with respect to
$H$, since both $H$ and $q$ are functions of time only and related to each other by (\ref{q}). That is, using
\begin{eqnarray}
&&\dot{q}=-H^2(1+q)q',\nonumber\\
&&\nonumber\\
&&\ddot{q}=H^4(1+q)^2q''+2H^3(1+q)^2q'+H^4(1+q)q'^2,
\end{eqnarray}
where the prime denotes derivatives with respect to $H$, we can bring (\ref{Hdot1}) into the following second-order
nonlinear differential equation for $q$ in $H$:
\begin{eqnarray}
&&\left[1+\frac{4H^2 \kappa \sigma_2} {D-2}\right](q+1)+\frac{H^2 \kappa \sigma_1}{(D-2) D}\Biggl\{8 - 2 D^2 + 6 D q^3 - H (D-9) D q'+H^2 D q'^2 + H^2 D q''\nonumber\\
&&~~~+D q^2 (20 - 4 D + 8 H q' + H^2 q'')+q [8 + 15 D - 6 D^2 - H (D-17) D q' +
H^2 D q'^2 + 2 H^2 D q'']\Biggr\}=0.\label{qeqn}\nonumber\\
\end{eqnarray}
Once the initial values $H_0$, $q_0$, and $q'_0$ are given, this equation can be numerically solved for the evolution of the deceleration parameter $q$ in $H$. This kind of analysis was previously exploited in the context of $f(R)$ gravity in \cite{Car-etal,DBD,PDG}. In the following analysis we will take the present-day values of the cosmological parameters $H$, $q$, $j$, and $s$ as \cite{Aghanim-etal,LQCX}
\begin{equation}
H_0=67.4~\text{km}\cdot \text{s}^{-1}\cdot \text{Mpc}^{-1}, ~~~~q_0=-0.71, ~~~~j_0=1.26,~~~~s_0=0.04.\label{Hqjs}
\end{equation}

Here, we shall consider two particular cases regarding our parameter
space $(\sigma_1, \sigma_2)$ corresponding to $\sigma_2=0$ (Case I) and $\sigma_1=0$ (Case II).\\
\\
\noindent\textbf{Case I:} When $\sigma_2$=0, the deceleration equation (\ref{qeqn}) becomes
\begin{eqnarray}
&&(q+1)+\frac{h^2 \kappa \sigma_1}{(D-2) D}\Biggl\{8 - 2 D^2 + 6 D q^3 - h (D-9) D q'+h^2 D q'^2 + h^2 D q''\nonumber\\
&&~~~+D q^2 (20 - 4 D + 8 h q' + h^2 q'')+q [8 + 15 D - 6 D^2 - h (D-17) D q' +
h^2 D q'^2 + 2 h^2 D q'']\Biggr\}=0.\label{qheqn}\nonumber\\
\end{eqnarray}
Note that in writing this equation we have rescaled the Hubble parameter $H$ in (\ref{qeqn}) as
\begin{equation}
H\equiv(100~\text{km}\cdot \text{s}^{-1}\cdot \text{Mpc}^{-1})~h
\end{equation}
to construct a dimensionless parameter $h$ and assumed all the numerical constants are absorbed into the constant $\sigma_1$.

\vspace{0.2cm}
\noindent
{\bf Remark 6}: Before presenting the solution of (\ref{qheqn}), it would be useful to look at the energy density $\rho_g$ related to the geometry: Since it must be positive, we can determine the sign of the constant $\sigma_1$ by evaluating the energy density at the present time. From  (\ref{rg1}), we can infer that the energy density at the present epoch is
\begin{equation}
\rho_{g}=\frac{(D-1)h_0^2 \sigma_1}{2 D} \left[2D(D-3)q_0+(D^2-4)+Dq_0^2-2Dj_0\right].
\end{equation}

\begin{figure}
\centering
\includegraphics[scale=0.96]{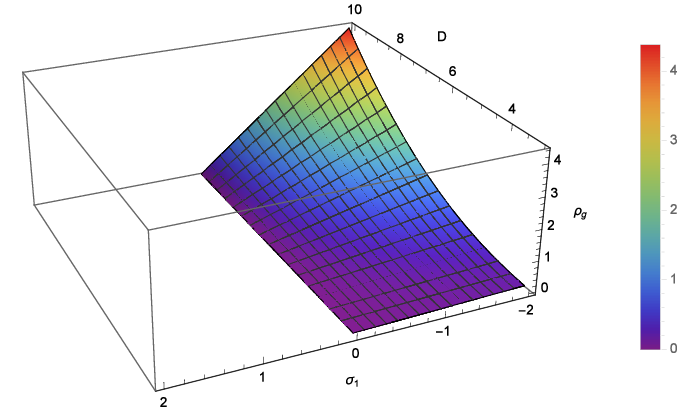}
\caption{The plot of $\rho_g$ vs $\sigma_1$ and $D$ for $h_0=0.674$, $q_0=-0.71$, and $j_0=1.26$.}
\label{fig:rg-D}
\end{figure}

\vspace{0.2cm}
\noindent
{\bf Remark 7}: Inserting $h_0$, $q_0$, and $j_0$ from (\ref{Hqjs}) into $\rho_g$ and graphing with respect to $\sigma_1$ and $D$, we obtain the plot shown in Fig. \ref{fig:rg-D}.
As is obvious from the figure, to have a positive $\rho_g$, the parameter $\sigma_1$ must be negative for all dimensions up to $D=11$.

\begin{figure}
\centering
\includegraphics[scale=0.96]{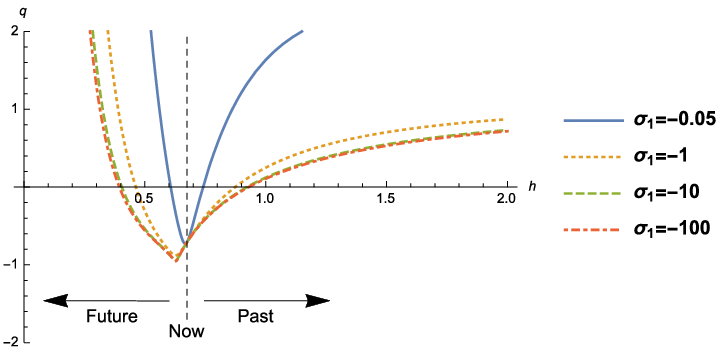}
\caption{The plot of $q$ as a function of $h$ for different values of $\sigma_1$ and for $D=4$, $\kappa=1$, $\sigma_2=0$, $h_0=0.674$, $q_0=-0.71$, and $q_0'=4.92$.}
\label{fig:q-h}
\end{figure}

Now we can solve the equation (\ref{qheqn}) numerically: First, to observe the effect of the value of $\sigma_1$ on the solution, we plot the solution of $q(h)$ for $D=4$ and for different values of $\sigma_1$, this is shown in Fig \ref{fig:q-h}.

\vspace{0.2cm}
\noindent
{\bf Remark 8}: Since $h$ is related to the inverse of the cosmic time $t$, in Fig.\ref{fig:q-h}, $h>h_0$ defines the past and $h<h_0$ defines the future in the cosmic evolution of the Universe. As is obvious from the figure, the deceleration parameter $q$ experiences two signature changes: one is at the past and the other is at the future.

\vspace{0.2cm}
\noindent
 {\bf Remark 9}: There is some time interval in which the deceleration parameter is negative (i.e. $q<0$). In particular, close to the present value of the Hubble parameter $h_0=0.674$, the deceleration parameter $q$ is negative. This means that the observed accelerated expansion phase of the Universe can be driven solely by the curvature of the spacetime.

 \vspace{0.2cm}
\noindent
{\bf Remark 10}: It can also be observed from Fig.\ref{fig:q-h} that, as the value of $\sigma_1$ decreases in negative (or increases in magnitude), the curves are opening out and approaching each other, and for very small values (or large magnitudes) they are becoming almost identical. This stems from the fact that, as the magnitude of the parameter $\sigma_1$ increases in (\ref{qheqn}), the first term in the equation can be neglected and the equation is effectively reduced to the one in which the curly bracketed expression equals zero.

\begin{figure}
\centering
\includegraphics[scale=0.96]{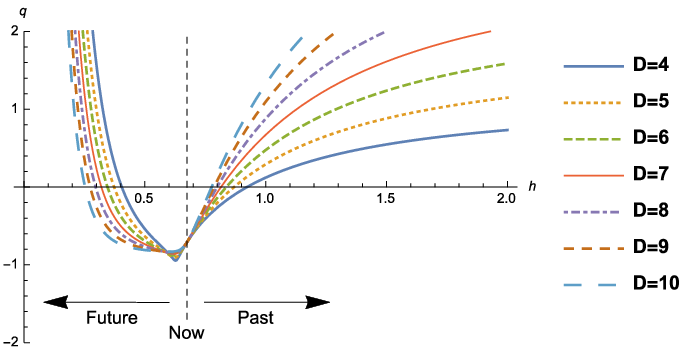}
\caption{Behavior of $q$ for $4\leq D\leq 10$, $\kappa=1$, $\sigma_1=-10$, $\sigma_2=0$, $h_0=0.674$, $q_0=-0.71$, $q'_0=4.92$.}
\label{fig:q-D}
\end{figure}
%

\vspace{0.2cm}
\noindent
{\bf Remark 11}: Additionally, we can also investigate the behavior of $q(h)$ with $D$. This is shown in Fig. \ref{fig:q-D}. Here it is explicitly seen that, as $D$ increases, the signature change of $q$ occurs at later times.

\begin{figure}
\centering
\includegraphics[scale=0.96]{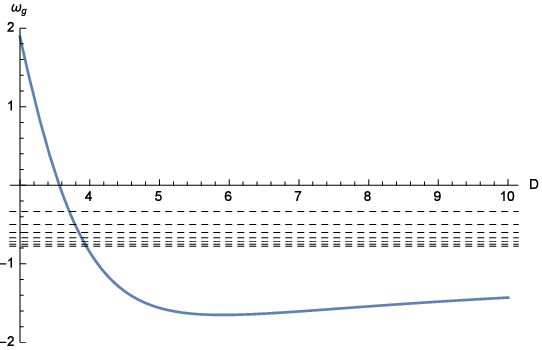}
\caption{Behavior of $w_g$ for $\sigma_2=0$, $q_0=-0.71$, $j_0=1.26$, $s_0=0.04$.}
\label{fig:wg-D}
\end{figure}

\vspace{0.2cm}
\noindent
On the other hand, one can also study the behavior of the equation of state parameter of the dark fluid $w_g$ stemmed from the terms proportional to $\sigma_1$ in
(\ref{rhoeff}) and (\ref{preseff}); that is, since $w_g=p_g/\rho_g$, at the present time $t_0$ we obtain
\begin{equation}
w_{g}=\frac{(D-2)(D+2)(D-5)-3D(D-3)q_0^2+2(D^3-9D^2+12D+8)q_0-2Ds_0-4(D-3)Dj_0}{(D-1)[-2D(D-3)q_0-(D^2-4)-Dq_0^2+2Dj_0]},
\end{equation}
the plot of which is given in Fig. \ref{fig:wg-D} with respect to $D$, where the dotted lines represent the upper bounds in (\ref{weff:cond}) decreasing in value as $D$ increases.\\
{\bf Remark 12}: As is obvious, $w_g$ is negative for all integers $4\leq D\leq 10$, consistently with Fig. \ref{fig:q-D}, and satisfying the condition (\ref{weff:cond}) for all dimensions. Specifically, for $D=4$, $w_g=-0.85$.

\vspace{0.1cm}

\noindent
{\bf Case II:} When $\sigma_1$=0 and $D\neq4$, the deceleration equation (\ref{qeqn}) becomes
\begin{equation}
\left[1+\frac{4H^2 \kappa \sigma_2} {D-2}\right](q+1)=0.
\end{equation}
Recalling that  $\sigma_2<0$ follows from the positiveness of $\rho_{g2}$, there are two possibilities:\\
\\
$(i)$~~  $\left[1+\frac{4H^2 \kappa \sigma_2} {D-2}\right]\neq 0$ and $q=-1$,
independently of the number of dimensions. Thus the deceleration parameter is always negative,  representing an accelerating Universe driven by the curvature of the spacetime. Also, one can show that for the equation of state of the geometric fluid $w_g$ stemmed from the terms proportional to $\sigma_2$ in (\ref{rhoeff}) and (\ref{preseff}) is
\begin{equation}
w_{g}=-\frac{D-5-4q}{D-1}=-1,
\end{equation}
for $q=-1$ independently of the number of dimensions. Hence, the geometric fluid can
derive the accelerating expansion of the Universe playing the role of an
effective cosmological constant.\\
\\
 $(ii)$ When $H^2=-\frac{(D-2)}{4\kappa \sigma_2}$, it follows that  $q=-1$ automatically.
This case represents an exact exponential solution for the scale factor, i.e $a(t)=a_0
e^{\pm\lambda t}$ where $\lambda= \sqrt{-(D-2)/4\sigma_2}$.
\newpage
\section{Some Particular Cosmological Solutions in Quadratic Gravity}
In this section, we shall investigate some particular solutions to the general equations  presented  in (\ref{reff}) and (\ref{peff}).

\subsection{Solutions with $\sigma_2=0$}
When $\sigma_2$ vanishes, the field equations (\ref{reff}) and (\ref{peff}) become
\begin{eqnarray}
&&\frac{ 1}{\kappa }\left[\frac{(D-1) (D-2)}{2} \left(\frac{k}{a^2}+\frac{\dot a^2}{a^2}\right)-\Lambda \right]\nonumber\\
&&~~~~~~~~~~~~~=\rho-\frac{(D-1)\sigma_1}{2 D a^4}\Biggl[k^2 (D-2)^2+\dot a^2 \left(2D (D-3)a \ddot a-4 k (D-2)\right)\nonumber\\
&&~~~~~~~~~~~~~~~~~~~~~~~~~~~~~~~~~~~~~~~~~~~~~~~~~~~~~~~~~~~~~~-\left(D^2-4\right)\dot a^4-D a^2 \ddot
a^2+2 D a^2 \dot a \dddot a\Biggr],\label{rhoD}\\
&&-\frac{1}{\kappa }\left[\frac{(D-2)(D-3)}{2}  \left(\frac{k}{a^2}+\frac{\dot
a^2}{a^2}\right)+(D-2) \frac{\ddot a}{a}-\Lambda   \right]\nonumber\\
&&~~~~~~=p+\frac{\sigma_1}{2 D a^4}\Biggl\{(D-2) (D-5) (D+2) \dot a^4-k^2 (D-2)^2 (D-5)  +a \ddot
a \left[8 k (D-2)-3 D(D-3)a \ddot a\right]\nonumber\\
&&~~~~~~~~~~~+2 \dot a^2 \left[2 k (D-2) (D-5)-\left(D ((D-9) D+12)+8\right) a \ddot a\right] -2 D \ddddot a a^3-4 (D-3)D \dddot a a^2 \dot a\Biggr\}.\label{pD}
\end{eqnarray}
It must be noted that these equations are valid either when $D=4$ and $(D-1) (D-2) \alpha+ D(D-3)\gamma\neq0$, or when $D\neq4$ and $(D-1) (D-2) \alpha+ D(D-3)\gamma=0$, or
when $D=4$ and $(D-1) (D-2) \alpha+ D(D-3)\gamma=0$. In four dimensions, these reduce to
\begin{eqnarray}
&&\frac{1}{\kappa a^4}\left[3 a^2 (\dot a)^2+3 k a^2-\Lambda a^4 \right]=\rho-\frac{3\sigma_1}{2a^4}\left[ 2 a^2 \dot a \dddot a- a^2 (\ddot
a)^2 +2 a (\dot a)^2\,\ddot a  -3 (\dot a)^4-2 k (\dot a)^2+ k^2  \right],\label{rhoeff1}\\
&&-\frac{1}{\kappa a^4}\left[2 a^3 \ddot
a+a^2 (\dot a)^2+k a^2-\Lambda a^4   \right]\nonumber\\
&&~~~~~~~~~~~~~~=p+\frac{\sigma_1}{2a^4}\left[ 2 a^3 \ddddot a+4 a^2 \dot a \dddot a+3 a^2 (\ddot a)^2 -12 a (\dot a)^2 \ddot a+3 (\dot a)^4-4 k   a \ddot a+2 k  (\dot a)^2 -k^2
\right].\label{peff1}
\end{eqnarray}
where $\sigma_1=4(3\alpha+ \beta$). This case is known as the critical gravity where (\ref{rhoeff1}) and (\ref{peff1}) reduces to the ones in the usual Einstein's gravity
when $\sigma_1=0$ \cite{GT, DT}. Here we have the following immediate Remarks:

\vspace{0.5cm}
\noindent
{\bf Remark 13}: When the critical  parameter $\sigma_1=0$  \cite{LP, DT1} it is quite
interesting that for this special case the above energy density and pressure expressions for the FLRW metric reduce to the corresponding expressions in pure Einstein theory.
This means that the highly nontrivial tensor field ${\mathcal E}_{\mu \nu}$ given in (\ref{Emunu}) reduces to
\begin{equation}
{\mathcal E}_{\mu \nu}=\alpha\left(2 R R_{\mu \nu}-\frac{1}{2} R^2 g_{\mu \nu}+\frac{1}{2} g_{\mu \nu} \square R+\nabla_{\mu} \nabla_{\nu} R -3\, \square R_{\mu \nu}-6 R_{\mu \rho \nu
\sigma} R^{\rho \sigma}+\frac{3}{2} g_{\mu \nu} R^{\rho \sigma} R_{\rho \sigma}\right),
\end{equation}
which vanishes identically for $\alpha\neq0$.

\vspace{0.5cm}
\noindent
{\bf Remark 14}: The case when $k=0$ and vanishing of the coefficient of $\sigma_1$ in (\ref{rhoeff1}) and (\ref{peff1}) corresponds to the work of Barrow et al. \cite{bar}. They
found a power law solution for $a$ and studied the stability of the solution.

\subsubsection{Exponential Solutions}

Now we shall study a special case which may correspond to the late-time accelerating era of the Universe. Let $a(t)=a_{0}\, e^{H_0 t}$, where $a_{0}$ and $H_0$ are the scale
factor and the Hubble constant at the time when the accelerating era begins, respectively. It is interesting that, for an exponentially expanding flat universe ($k=0$), the
contributions of the quadratic gravity terms related to $\sigma_1$ in (\ref{rhoD}) and (\ref{pD}) vanish identically. This means that the presence of the bare cosmological
constant $\Lambda$ is crucial for having exponential solutions in arbitrary $D$ dimensions. When $k\neq0$, the field equations (\ref{rhoD}) and (\ref{pD}) become
\begin{eqnarray}
\rho&=&\frac{(D-1)(D-2)}{2D\kappa a_{0}^4}\left[D\beta_{0}\, a_{0}^4+ k \,a_{0}^2 \beta_{2}\, e^{-2 H_0 t}+(D-2) k^2 \kappa \sigma_1 e^{-4H_0 t}\right], \\
p&=& -\frac{D-2}{2D\kappa a_{0}^4}\left[D\beta_{0}\, a_{0}^4+ (D-3)k \, a_{0}^2\, \beta_{2}\, e^{-2 H_0 t}+(D-2)(D-5) k^2 \kappa \sigma_1 e^{-4H_0 t}\right],
\end{eqnarray}
where $\beta_{0}=H_0^2-\frac{2\Lambda}{(D-1)(D-2)}$, $\beta_{2}=D -4 \kappa\sigma_1H_0^2$, and $\sigma_1=4(D-1)\alpha+D\beta$. We have the following consequences.

\vspace{0.5cm}
\noindent
(a) If $\beta_{0}>0$ and $\beta_{2}>0$, the energy density remains positive for all $t$. At late times when $t \to \infty$, we have $\rho\to\frac{(D-1)(D-2)\beta_{0}}{2\kappa}$ and
$p\to -\frac{(D-2)\beta_{0}}{2\kappa}$. Hence, the equation of state is  of dark energy type, i.e $p\to-\frac{\rho}{D-1}$.

\vspace{0.5cm}
\noindent
(b) When $\beta_{0}=0$ or $H_0=\sqrt{\frac{2\Lambda}{(D-1)(D-2)}}$ and as $t \to \infty$, then we obtain
\begin{eqnarray}
\rho&=&\frac{(D-1)(D-2)}{2D\kappa a_{0}^4}\left[k \,a_{0}^2 \beta_{2}\, e^{-2 H_0 t}+(D-2) k^2 \kappa \sigma_1 e^{-4H_0 t}\right], \\
p&=& -\frac{D-2}{2D\kappa a_{0}^4}\left[ (D-3)k \, a_{0}^2\, \beta_{2}\, e^{-2 H_0 t}+(D-2)(D-5) k^2 \kappa \sigma_1 e^{-4H_0 t}\right].
\end{eqnarray}
In these expressions, since the last terms decay faster than the first terms, one can deduce an equation of state at late times
\begin{equation}
p=-\frac{D-3}{D-1}\rho.
\end{equation}
It can be seen that, for $D\geq4$ and $\rho>0$, the pressure is always negative. In $D=4$, this gives $p=-\rho/3$ which corresponds to the equation of state of cosmic strings \cite{vilen}.

\vspace{0.5cm}
\noindent
(c) To have positive pressure, we let $\beta_{0}=0$ and $\beta_{2}=0$ together so that $H_0=\sqrt{\frac{2\Lambda}{(D-1)(D-2)}}$ and $4 \kappa\sigma_1H_0^2=D$. We find
\begin{eqnarray}
\rho&=&\frac{(D-1)(D-2)^2}{2D a_{0}^4}k^2\sigma_1 e^{-4H_0 t}, \\
p&=& -\frac{(D-2)^2(D-5)}{2D a_{0}^4} k^2  \sigma_1 e^{-4H_0 t}.
\end{eqnarray}
This special solution can only be obtained when both $k$ and $\sigma_1$ are nonvanishing. Hence it can be obtained neither in Einstein theory nor in the work of Barrow et al \cite{bar}. In other words, it is the effect of the quadratic gravity which predicts a de Sitter era at late times with the acceleration of the expansion being constant,
i.e. the square of the Hubble constant ($H_0=\frac{\dot a}{a}$) at the beginning of the accelerating phase. Furthermore, the above expressions for the energy density and
pressure of the fluid provides an equation of state
\begin{equation}
p=-\frac{D-5}{D-1}\rho,
\end{equation}
which, for $D\geq6$ and $\rho>0$, the pressure is always negative. It can also be seen that in $D=5$ the pressure vanishes corresponding to the equation of state of dust
matter, and in $D=4$ it becomes $p=\frac{1}{3} \rho$ which corresponds to the equation of state of radiation. This solution is valid for both closed ($k=1$) and open ($k=-1$)
universes.

At this point, it should be stressed that all the above equations of state can mimic a variety of sources of ``geometric'' origin filling the Universe which accelerates like
the de Sitter universe at late times.

\subsubsection{A More General Solution}

Let us assume that $a(t)$ satisfies the following differential equation
\begin{equation}
\dot{a}^2=f(a),
\end{equation}
where $f(a)$ is any arbitrary function of the scale factor $a(t)$. Hence, the energy density and pressure expressions in (\ref{rhoD}) and (\ref{pD}) become
\begin{eqnarray}
&&\rho=\frac{(D-1)(D-2)}{2\kappa a^4}\Biggl\{(f+k)a^2-\frac{2\Lambda}{(D-1)(D-2)}a^4\nonumber\\
&&~~~~~~+\frac{\sigma_1\kappa}{D}\left[(D-2)k^2-(D+2)f^2-\frac{D}{4(D-2)}a^2f'^2-4kf+\frac{D(D-3)}{(D-2)}aff'+\frac{D}{(D-2)}a^2ff''\right]\Biggr\}, ~~~~~~\\
&&p=-\frac{(D-1)(D-2)}{2\kappa a^4}\Biggl\{\frac{D-3}{D-1}(f+k)a^2-\frac{2\Lambda}{(D-1)(D-2)}a^4+\frac{1}{(D-1)}a^3f'\nonumber\\
&&~~~~~~-\frac{\sigma_1\kappa}{D(D-1)}\Biggl[(D+2)(D-5)f^2-(D-2)(D-5)k^2+4kaf'
-\frac{3D(D-3)}{4(D-2)}a^2f'^2+4(D-5)kf\nonumber\\
&&~~~~~~~~~~~~~~~~~~~~~~-\frac{D^3-9D^2+12D+8}{(D-2)}aff'-\frac{2D(D-3)}{(D-2)}a^2ff''-\frac{D}{2(D-2)}a^3(f'f''+2ff''')\Biggr]\Biggr\},
\end{eqnarray}
where $f^{\prime}=\frac{df}{da}$. Now we shall give some examples:

\vspace{0.5cm}
\noindent
\textbf{Example 1}: Let $f=a_0k+\frac{a_{1}}{a}+a_3a^2$. For this choice, the acceleration of the Universe, $\ddot{a}=f'/2$, becomes positive when $a >
(\frac{a_{1}}{2a_3})^{1/3}$. Taking $a_0=-1$ and $a_3=\frac{2\Lambda}{(D-1)(D-2)}$, we get
\begin{eqnarray}
\rho&=&\frac{2(D-4)\sigma_1\Lambda^2}{D(D-2)^2}-\frac{a_1(D-1)}{2D\kappa a^6}\Biggl\{D(D-2)a^3\nonumber\\
&&+\frac{\kappa\sigma_1}{4}\biggl[(8D^2-19D-16)a_1+12D(D-3)ka-\frac{8(D+2)(D-4)\Lambda}{(D-1)(D-2)}a^3\biggr]\Biggr\}, \\
p&=&-\frac{2(D-4)\sigma_1\Lambda^2}{D(D-2)^2}-\frac{a_1}{2D\kappa a^6}\Biggl\{D(D-2)(D-4)a^3\nonumber\\
&&+\frac{\kappa\sigma_1}{4}\biggl[(D-7)(8D^2-19D-16)a_1+12D(D-3)(D-6)ka-\frac{8(D+2)(D-4)\Lambda}{(D-1)(D-2)}a^3\biggr]\Biggr\},
\end{eqnarray}
where $a_{1}$ is an arbitrary constant. From these expressions, it can readily be observed that, when $a_1=0$, the equation of state of the fluid reduces to the form
\begin{equation}
p=-\rho,
\end{equation}
which corresponds to a cosmological constant equation of state for $D\neq4$. For $D=4$ and $a_1=0$, the energy density and pressure vanish identically. This means that, for
$f=-k+\frac{\Lambda}{3}\,R^2$, the FLRW metric is the vacuum solution of the quadratic gravity field equations. In particular, when $k=0$, this is the usual de Sitter
solution with the scale factor $a=a_0e^{\sqrt{\Lambda/3}t}$.

\vspace{0.5cm}
\noindent
\textbf{Example 2}: Let $f=a_0k+\frac{a_{2}}{a^2}+a_3a^2$. This time, the acceleration of the Universe, $\ddot{a}=f'/2$, becomes positive when $a >
(\frac{a_{2}}{a_3})^{1/4}$. Taking $a_0=-1$, we obtain
\begin{eqnarray}
&&\rho=\frac{D-1}{2D\kappa}\left[(D-1)(D-4)\kappa\sigma_1a_3^2+D(D-2)\left(a_3-\frac{2\Lambda}{(D-1)(D-2)}\right)\right]\nonumber\\
&&~~~~~~+\frac{a_2(D-1)}{2D\kappa a^8}\Biggl\{D(D-2)a^4-\kappa\sigma_1\biggl[(D-4)(1+3D)a_2-4D(D-4)ka^2+2(D^2-5D-4)a_3a^4\biggr]\Biggr\},~~~~~~ \\
&&p=-\frac{D-1}{2D\kappa}\left[(D-1)(D-4)\kappa\sigma_1a_3^2+D(D-2)\left(a_3-\frac{2\Lambda}{(D-1)(D-2)}\right)\right]\nonumber\\
&&~~~~~~-\frac{a_2}{2D\kappa a^8}\Biggl\{D(D-2)(D-5)a^4-\kappa\sigma_1\biggl[(D-4)(D-9)(1+3D)a_2-4D(D-4)(D-7)ka^2\nonumber\\
&&~~~~~~~~~~~~~~~~~~~~~~~~~~~~~~~~~~~~~~~~~~~~~~~~~~~~~~~~~~~~~~~~~~~~~~~~~~~~~~~~~~~~~~~~+2(D-5)(D^2-5D-4)a_3a^4\biggr]\Biggr\},
\end{eqnarray}
where $a_{2}$ and $a_{3}$ are arbitrary constants. Here we have the following special case: When $D=4$,
\begin{eqnarray}
\rho&=& \frac{3 a_{3}-\Lambda}{\kappa}+\frac{3 a_{2}}{\kappa a^4} (1+2\kappa\sigma_1 a_{3}), \\
p&=&-\frac{3 a_{3}-\Lambda}{\kappa}+\frac{ a_{2}}{\kappa a^4} (1+2\kappa\sigma_1a_3).
\end{eqnarray}
Now choosing $a_3=\Lambda/3$, the equation of state becomes
\begin{equation}
p=\frac{1}{3}\rho,
\end{equation}
which corresponds to an radiation equation of state for arbitrary $a_2$. This means that, for $f=-k+\frac{a_{2}}{a^2}+a_3a^2$, the spatially flat ($k=0$) FLRW metric solves the quadratic gravity field equations with the scale factor
\begin{equation}
a(t)=\frac{1}{\sqrt{2a_3}}e^{\sqrt{a_3}t} \sqrt{1-a_2 a_3e^{-4\sqrt{a_3}t}}.
\end{equation}
As it can be seen, $a(t)\to\frac{1}{\sqrt{2a_3}}e^{\sqrt{a_3}t}$ as $t\to\infty$, which means that, although the equation of state represents a radiation field, there is a de
Sitter like expansions at late times.

\vspace{0.5cm}
\noindent
\textbf{Example 3}: Let $f=a_0e^{ma}$ where $a_0$ and $m$ are two real arbitrary constants. For this function, the scale factor is
\begin{equation}
a=-\frac{2}{m}\log\left(1-\frac{m\sqrt{a_0}}{2}t\right)
\end{equation}
where $a>0$ requires $m<0$. Thus, this corresponds to a decelerating universe. In this case, the energy density and pressure take the form
\begin{eqnarray}
\rho&=&\frac{a_0^2 (D-1) \sigma_1  \left[3 m^2 D a^2+4 m (D-3) D a-4 \left(D^2-4\right)\right]}{8 D a^4}e^{2 m a}+\frac{a_0 (D-1)(D-2) }{2 \kappa  a^2}e^{m a}-\frac{\Lambda
}{\kappa }, \\
p&=&-\frac{a_0^2 \sigma_1 \left[6 m^3 D a^3+11 m^2 (D-3) D a^2+4 m \left(D^3-9 D^2+12 D+8\right) a-4 (D-5) (D-2) (D+2)\right]}{8 D a^4}e^{2 m a}\nonumber\\
&&-\frac{a_0 (D-2) (m a+D-3)}{2 \kappa  a^2}e^{m a}+\frac{\Lambda }{\kappa }.
\end{eqnarray}

\vspace{0.5cm}
\noindent
\textbf{Example 4}: Let $f=a_0a^m$ where $a_0$ and $m$ are two arbitrary real constants. This function produces the scale factor as
\begin{equation}
a=\left[\left(\frac{2-m}{2}\right)\sqrt{a_0}\right]^{\frac{2}{2-m}}t^{\frac{2}{2-m}}
\end{equation}
where $a_0>0$ and $m\neq2$. From this scale factor, one obtains the acceleration as $\ddot{a}=\frac{a_0m}{2}a^{m-1}$. To assure the positiveness of $a$ and $\ddot{a}$, it
must be that $0<m<2$. In this case, the energy density and pressure are
\begin{eqnarray}
&&\rho=\frac{1}{\kappa}\left[\frac{(D-2) (D-1) \left(a_0 a^m+k\right)}{2 a^2}-\Lambda\right]\nonumber\\
&&~~~~~~+\frac{(D-1) \sigma_1}{8 D a^4} \Biggl\{a_0^2 \left[3 m^2 D+4 m (D-4) D-4 (D-2)(D+2)\right] a^{2 m}\nonumber\\
&&~~~~~~~~~~~~~~~~~~~~~~~~~~~~~~~~~~~~~~~~~~~~~~~~~~~~~~~~~~~~~~~-16 a_0 k (D-2) a^m+4 k^2 (D-2)^2\Biggr\}, \\
&&p=-\frac{1}{\kappa}\left[\frac{1}{2} a_0 (D-2) (m+D-3) a^{m-2}+\frac{k (D-1)(D-5)}{2 a^2}-\Lambda\right]\nonumber\\
&&-\frac{\sigma_1}{8 D a^4} \Biggl\{a_0^2 \left[6 m^3 D+m^2 D (11 D-47)+4 m \left(D^3-11 D^2+20 D+8\right)-4 (D+2)(D-2)(D-5)\right] a^{2 m}\nonumber\\
&&~~~~~~~~~~~~~~~~~~~~~~~~~~~~~~~~~~~~~~~~~~-16 a_0 k (D-2) (m+D-5) a^m+4 k^2 (D-5) (D-2)^2\Biggr\}.
\end{eqnarray}
It can be observed that, when $t\to\infty$, the contributions of the higher curvature terms decay faster than the Einstein terms, and hence
\begin{eqnarray}
\rho&\to&\frac{1}{\kappa}\left[\frac{(D-2) (D-1) \left(a_0 a^m+k\right)}{2 a^2}-\Lambda\right]\nonumber\\
p&\to&-\frac{1}{\kappa}\left[\frac{1}{2} a_0 (D-2) (m+D-3) a^{m-2}+\frac{k (D-1)(D-5)}{2 a^2}-\Lambda\right].\nonumber
\end{eqnarray}
Now, for the flat universe and $\Lambda=0$, we can deduce that
\begin{equation}
w\equiv\frac{p}{\rho}=-\frac{D-3+m}{D-1}
\end{equation}
which reduces to $w=-\frac{1}{3}(1+m)$ in $D=4$. Since $0<m<2$, it must be that $-1<w<-\frac{D-3}{D-1}$. One can see that as $D$ increases, the range squeezes and the upper
bound approaches to -1.

On the other hand, at early times (as $t\to0$), the only conditions on the parameters of the scale factor are $a_0>0$ and $m<2$. There are the following cases:
\begin{itemize}
\item For $m<0$, we have
\begin{eqnarray}
\rho&\to&\frac{(D-1) \sigma_1a_0^2a^{2 m}}{8 D a^4}\left[3 m^2 D+4 m (D-4) D-4 (D-2)(D+2)\right], \\
p&\to&-\frac{(D-5+2m)\sigma_1a_0^2a^{2 m}}{8 D a^4}\left[3 m^2 D+4 m (D-4) D-4 (D-2)(D+2)\right].
\end{eqnarray}
Then one can deduces the following equation of state
\begin{equation}
w\equiv\frac{p}{\rho}=-\frac{D-5+2m}{D-1},
\end{equation}
which becomes $w=\frac{1-2m}{3}>\frac{1}{3}$ in $D=4$. For $m=-1$, it gives the stiff fluid equation of state, $w=1$.
\item For $m=0$, we have
\begin{eqnarray}
\rho&\to&\frac{(D-1) (D-2) \sigma_1}{2 D a^4} \left[k^2 (D-2)-4 a_0 k -(D+2)a_0^2\right], \\
p&\to&-\frac{\sigma_1(D-2) (D-5)}{2 D a^4} \left[k^2 (D-2)-4 a_0 k - (D+2)a_0^2\right].
\end{eqnarray}
Then
\begin{equation}
w\equiv\frac{p}{\rho}=-\frac{D-5}{D-1}.
\end{equation}
This becomes $w=\frac{1}{3}$ in $D=4$ and $w=0$ in $D=5$. The former represents the radiation and the latter represents the dust matter.
\item For $0<m<2$, we have
\begin{eqnarray}
\rho&\to&\frac{(D-1)(D-2)^2 \sigma_1k^2}{2 D a^4}\\
p&\to&-\frac{(D-2)^2(D-5) \sigma_1k^2}{2 D a^4}.
\end{eqnarray}
Then
\begin{equation}
w\equiv\frac{p}{\rho}=-\frac{D-5}{D-1}.
\end{equation}
Again this represents the radiation ($w=\frac{1}{3}$) in $D=4$ and the dust matter ($w=0$) in $D=5$.
\end{itemize}


\subsubsection{Approximate Solutions Near Criticality with $\sigma_2=0$ }

Now, we assume that, in generic gravity theories in (\ref{action}), the action contains some number of coupling constants and hence $A$ and $B$ in (\ref{rho}) and
(\ref{pres}) are functions of these coupling constants and/or some combinations $\alpha_{i}$ of them. Assuming that these coupling constants are relatively smaller than the
other parameters in these functions, we can expand the scale factor $a$ in terms of these parameters
\begin{equation}\label{aalpha}
a(t,\alpha_i)=a_0(t)+\sum_i\alpha_i a_{i}(t)+\mathcal{O}(\alpha_i^2).
\end{equation}
Following this approach, we obtain the functions $A$ and $B$ as
\begin{eqnarray}
&&A=\sum_i\alpha_i A_{i}+\mathcal{O}(\alpha_i^2) , \\
&&B=\sum_i\alpha_i B_{i} +\mathcal{O}(\alpha_i^2),
\end{eqnarray}
where $A_i$ and $B_i$ are functions depending on the explicit gravity theory. In what follows we shall keep only the terms linear in $\alpha_i$. Hence Eqs.(\ref{rho}) and
(\ref{pres}) reduce to
\begin{eqnarray}
&&\rho-\frac{1}{\kappa}\left[\frac{(D-1)(D-2)}{2}\rho_{10}-\Lambda\right] +\sum_i\alpha_i\left[A_{i}-B_{i}-\frac{(D-1)(D-2)}{2\kappa}\rho_{1i}\right]=0,\label{ralpha}\\
&&p+\frac{1}{\kappa}\left[\frac{(D-1)(D-2)}{2}\rho_{10}-(D-2)\rho_{20}-\Lambda\right]
-\sum_i\alpha_i\left[A_{i}-\frac{(D-2)}{\kappa}\left(\frac{D-1}{2}\rho_{1i}-\rho_{2i}\right)\right]=0,~~~~~~ \label{palpha}
\end{eqnarray}
where
\begin{eqnarray}
&&\rho_{10}=H_0^2 +\frac{k}{a_0^2},\label{}\\
&&\rho_{20}=H_0^2 +\frac{k}{a_0^2}-\frac{\ddot a_0}{a_0}=\dot{H_0}+\frac{k}{a_0^2},\label{}\\
&&\rho_{1i}= \frac{6}{a_0} \left(\dot a_{i}H_0- a_{i} H_0^2-\frac{k a_{i}}{a_0^2}, \right)\\
&&\rho_{2i}=\frac{2}{a_0} \left[\dot a_{i}H_0- a_{i} H_0^2-\frac{k a_{i}}{a_0^2}
+\frac{1}{2}\left(\frac{\ddot a_0 a_i}{a_0} +\ddot a_i\right)\right],
\end{eqnarray}
where $H_0=\dot a_0/a_0$. Now one can find $a_{i}$ from (\ref{ralpha}) by assuming that the extra terms in $\alpha_i$ vanish, i.e.,
\begin{equation}\label{cond}
\sum_i\alpha_i\left[A_{i}-B_{i}-\frac{(D-1)(D-2)}{2\kappa}\rho_{1i}\right]=0.
\end{equation}
{\bf Remark 15}: Then the expression for $\rho$ becomes exactly the energy density without the modification of the  generic gravity, but the expression for $p$ has contributions from the
generic gravity theory.


Now we will specifically consider the critical quadratic gravity theory given by the action (\ref{eq:Quadratic_action}) with $\sigma_2=0$ in (\ref{s22}). Taking $i=1$ and
$\alpha_i=\sigma_1$ in the above formulation, from (\ref{aalpha}) the scale factor becomes
\begin{equation}\label{asigma}
a(t,\sigma_1)=a_0(t)+\sigma_1 a_{1}(t)+\mathcal{O}(\sigma_1^2).
\end{equation}
and the field equation (\ref{ralpha}) reads as
\begin{eqnarray}
&&\rho-\frac{1}{\kappa}\left[\frac{(D-1)(D-2)}{2}\left(H_0^2 +\frac{k}{a_0^2}\right)-\Lambda\right] -\frac{(D-1)(D-2)\sigma_1}{\kappa a_0}\Biggl\{\dot a_{1}H_0- a_{1}
H_0^2-\frac{k a_{1}}{a_0^2}\nonumber\\
&&~~~~~~~~~~~~~~~~~~~~+\frac{\kappa}{2 D (D-2)a_0^3}\Biggl[k^2 (D-2)^2+\dot a_0^2 \left[2D (D-3)a_0 \ddot a_0-4 k (D-2)\right]\nonumber\\
&&~~~~~~~~~~~~~~~~~~~~~~~~~~~~~~~~~~~~~~~~~~~~~~~~~~~~~~~~~-\left(D^2-4\right)\dot a_0^4-D a_0^2 \ddot
a_0^2+2 D a_0^2 \dot a_0 {\dddot a}_0\Biggr]\Biggr\}=0.
\end{eqnarray}
To determine $a_{1}$, we assume that the coefficient terms of $\sigma_1$ vanish, ie.,
\begin{eqnarray}\label{condg}
&&\dot a_{1}H_0- a_{1}\left( H_0^2+\frac{k}{a_0^2}\right)+\frac{\kappa}{2 D (D-2)a_0^3}\Biggl[k^2 (D-2)^2+\dot a_0^2 \left[2D (D-3)a_0 \ddot a_0-4 k (D-2)\right]\nonumber\\
&&~~~~~~~~~~~~~~~~~~~~~~~~~~~~~~~~~~~~~~~~~~~~~~~~~~~~~~~~~~~~~~~~~~~~-\left(D^2-4\right)\dot a_0^4-D a_0^2 \ddot
a_0^2+2 D a_0^2 \dot a_0 {\dddot a}_0\Biggr]=0,~~~~~~
\end{eqnarray}
which can also be obtained from (\ref{cond}). This equation can be rewritten in the following 1st order
linear differential equation form
\begin{equation}\label{diff}
\dot a_1 +R(t) a_1=S(t),
\end{equation}
where
\begin{eqnarray}
&&R(t)=-\left( H_0+\frac{k}{a_0^2H_0} \right),\\
&&S(t)=-\frac{\kappa}{2 D (D-2)a_0^2\dot{a_0}}\Biggl\{k^2 (D-2)^2+\dot a_0^2 \left[2D (D-3)a_0 \ddot a_0-4 k (D-2)\right]\nonumber\\
&&~~~~~~~~~~~~~~~~~~~~~~~~~~~~~~~~~~~~~~~~~~~~~~~~~~~~~~~~~~~~-\left(D^2-4\right)\dot a_0^4-D a_0^2 \ddot
a_0^2+2 D a_0^2 \dot a_0 {\dddot a}_0\Biggr\}.
\end{eqnarray}
The equation (\ref{diff}) admits the general solution
\begin{equation}\label{a1}
a_1(t)=  \frac{C}{\lambda(t)} +\frac{\mu(t)}{\lambda(t)},
\end{equation}
where $C$ is an integration constant and
\begin{equation}
\lambda(t)=e^{\int R(t) dt}, ~~~~~\mu(t)=\int \lambda(t)  S(t) dt.
\end{equation}
Taking $D=4$, we consider the following cases:
\begin{enumerate}
\item Letting $a_0(t)=b_{0} e^{c_{0} t}$, where $b_0$ and $c_0$ are constants, we find $a_{1}(t)$ as
\begin{equation}
a_1(t)=e^{c_0 t}\left[C e^{-d_0 e^{-2c_0 t}}+\frac{\kappa}{4b_0}\left(k
 e^{-2c_0t} -4c_0^2 b_0^2 \right) \right],  ~~~~~d_0=\frac{k}{2c_0^2 b_0^2}
\end{equation}
Hence $a(t, \sigma_1)$ in (\ref{asigma}) reads as
\begin{equation}
a(t, \sigma_1)=b_{0}e^{c_0 t}+\sigma_1 e^{c_0 t}\left[C e^{-d_0 e^{-2c_0 t}}+\frac{\kappa}{4b_0}\left(k
 e^{-2c_0t} -4c_0^2 b_0^2 \right) \right].
\end{equation}
For a flat universe $(k=0)$, it reduces to
\begin{equation}
a(t, \sigma_1)=\chi_0e^{c_0 t}, ~~~~\chi_0=b_0 +\sigma_1\left(C-\kappa c_0^2 b_0\right).
\end{equation}
Then, the contribution of the higher curvature terms in the scale factor
for a flat universe has the same exponential form $e^{c_0 t}$ as in GR. For $\chi_0>0$, the total
acceleration, i.e $\ddot a(t,\sigma_1)$,  is positive.
\item Letting $a_0(t)=m_{0} t^{n}$, where $m_0$ and $n$ are constants, we find $a_1(t)$
in (\ref{a1}) with
\begin{eqnarray}
&&\lambda(t)=t^{-n}\cdot  e^{b t^{2(1-n)}},\\
&&\mu(t)=\int e^{bt^{2(1-n)}}\left(-\frac{3}{4} m_0\kappa n(1-2n) t^{-3} +\frac{\kappa
nk}{2m_0} t^{-2n-1} -\frac{\kappa k^2}{4nm_0^3} t^{1-4n}\right)dt,
\end{eqnarray}
where $b=-\frac{k}{2nm_0^2 (1-n)}$. Here we consider the following two cases.
\begin{itemize}
\item $k=0, ~~n=\frac{1}{2}$ corresponding to the radiation era in a flat
universe in the context of GR.\ For this
case, we  obtain
\begin{equation}\label{}
a_1(t)= C t^{\frac{1}{2}},
\end{equation}
and then
\begin{equation}\label{}
a(t, \sigma_1)= \chi_0 t^{\frac{1}{2}}, ~~~\chi_0=m_0 +\sigma_1 C
\end{equation}
Then, the contribution of the higher curvature terms in the total scale factor
for a flat universe has the same power law form $t^{\frac{1}{2}}$ as in GR. In contrast to the previous case of the scale factor with an exponential form, here we see
that the total acceleration $\ddot a(t, \sigma_1)=-\frac{1}{4}
\chi_0 t^{-\frac{3}{2}}$ is positive if $\chi_0<0$. For $\chi_0>0$, the higher
curvature modifications cannot support the acceleration of the universe.
\item $k=0, ~~n=\frac{2}{3}$ corresponding to the dust matter era in  a flat
universe in the context of GR. For this
case, we  obtain
\begin{equation}\label{}
a_1(t)= C t^{\frac{2}{3}}-\frac{1}{12}\kappa m_0 t^{-\frac{4}{3}},
\end{equation}
and then
\begin{equation}\label{}
a(t, \sigma_1)= \chi_0 t^{\frac{2}{3}}-\frac{1}{12}\kappa m_0 t^{-\frac{4}{3}}, ~~~\chi_0=m_0 +\sigma_1 C
\end{equation}
In contrast to two previous cases, here we see that the contribution of
the higher curvature terms in the scale factor are not of the same kind of  GR, and there is  an extra $t^{-\frac{4}{3}}$  type of contribution.
The total acceleration has the form $\ddot a(t, \sigma_1)=-\frac{2}{9}
\chi_0 t^{-\frac{4}{3}}-\frac{28}{108}\sigma_1 \kappa m_0 t^{-\frac{10}{3}}$
which is negative for $(m_0, C, \sigma_1, \kappa)>0$.
Then, the higher curvature modifications cannot support the acceleration of
the universe in a matter dominated era.\end{itemize}
\end{enumerate}


\subsection{Solutions with $\sigma_1=0$}

When $\sigma_1=0$ and $D\neq4$, the field equations (\ref{reff}) and (\ref{peff}) become
\begin{eqnarray}
&&\frac{ 1}{\kappa }\left[\frac{1}{2} (D-2) (D-1) \left(\frac{k}{a^2}+\frac{\dot
a^2}{a^2}\right)-\Lambda \right]=\rho-(D-1)\sigma_2\left(\frac{k}{a^2}+\frac{\dot
a^2}{a^2}\right)^2,\label{rs2D}\\
&&-\frac{1}{\kappa }\left[\frac{(D-2)(D-3)}{2}  \left(\frac{k}{a^2}+\frac{\dot
a^2}{a^2}\right)+(D-2) \frac{\ddot a}{a}-\Lambda   \right]\nonumber\\
&&~~~~~~~~~~~~~~~~~~~~~~~~~~=p+\sigma_2\left(\frac{k}{a^2}+\frac{\dot
a^2}{a^2}\right) \left[(D-5) \left(\frac{k}{a^2}+\frac{\dot
a^2}{a^2}\right)+4 \frac{\ddot a}{a}\right].\label{ps2D}
\end{eqnarray}
In this case, when $[(D-1) (D-2) \alpha+ D(D-3)\gamma]=0$ (i.e. $\sigma_2=0$), one again recovers the expressions in the Einstein gravity in $D\neq4$ dimensions.

To see the effects of quadratic gravity terms on the expansion of the universe at late times, neglecting $\rho$ and $p$ of baryonic matter in (\ref{rs2D}) and (\ref{ps2D}).
Then, from (\ref{rs2D}), one can immediately obtain the following solutions for the scale factor:
\begin{equation}
a(t)=
\begin{cases}
\frac{\sinh[\sqrt{h_0}(t-t_0)]}{\sqrt{h_0}}~~ \text{for $k=-1$},\\
\\
e^{\sqrt{h_0}(t-t_0)}~~ \text{for $k=0$},\\
\\
\frac{\cosh[\sqrt{h_0}(t-t_0)]}{\sqrt{h_0}}~~ \text{for $k=+1$},
\end{cases}
\end{equation}
where $t_0$ is an arbitrary integration constant and
\begin{equation}
h_0\equiv-\frac{D-2}{4\kappa\sigma_2}\left[1\mp \sqrt{1+\frac{16\sigma_2\Lambda}{(D-1)(D-2)^2}}\right]
\end{equation}
with $\sigma_2\Lambda\geq-\frac{(D-1)(D-2)^2}{16}$. One can verify that, with this solution, (\ref{ps2D}) is identically satisfied. The energy density and pressure of the
geometric fluid becomes
\begin{eqnarray}
\rho_{g2}&=&-\sigma_2(D-1)h_0^2, \\
p_{g2}&=&\sigma_2(D-1)h_0^2.
\end{eqnarray}
{\bf Remark 16}: These give an equation of state $p_{g2}=-\rho_{g2}$ which corresponds to the vacuum equation of state. It must be observed that the positiveness of the energy density requires
$\sigma_2<0$. Also it should be stated that in the absence of cosmological constant ($\Lambda=0$), the higher curvature terms in the theory behaves like en effective
cosmological constant driving the late time exponentially accelerating expansion.
These results are consistent with the ones discussed in Case II in section
4.

\section{Conclusion}
We have given all our findings both in abstract and in the introduction. Here we give a short summary of this work.
We consider FLRW cosmology in the context of generic gravity theories in which the action includes all the combinations of the metric tensor, curvature tensor, and covariant derivatives of the curvature tensor of any order. Very recently we showed that in such theories with FLRW geometry, contributions of all higher order terms reduce to a perfect fluid form which we call it now as the geometric fluid. Hence all generic theories of gravity in FLRW geometry are equivalent to Einstein's theory of general relativity where the source term contains both matter and geometric fluids. We propose that the source of dark energy/matter is this geometrical fluid arising from higher order gravity theories. Choosing any higher order gravity, the parameters of the theory can be suitably arranged that  the corresponding geometric fluid contributes to the accelerated expansion of the universe. We verified our assertion by taking the quadratic gravity as an example. Furthermore we have given some particular exact cosmological solutions of quadratic gravity theory with matter and geometrical fluids.


\end{document}